\documentclass[apjl]{emulateapj}

\slugcomment{Submitted to ApJ Letters}

\usepackage{natbib}

\begin{document}

\setlength\tabcolsep{4pt}
\tabletypesize{\scriptsize}

\title{Massive and Newly Dead: Discovery of a Significant Population
  of Galaxies with High Velocity Dispersions and Strong Balmer Lines
  at $z\sim1.5$ from Deep Keck Spectra and HST/WFC3 Imaging}

\author{Rachel Bezanson\altaffilmark{1}, Pieter van Dokkum\altaffilmark{1}, Jesse van de Sande \altaffilmark{2}, Marijn Franx\altaffilmark{2}, Mariska Kriek \altaffilmark{3}}

\altaffiltext{1}{Department of Astronomy, Yale University, New Haven, CT 06520-8101}
\altaffiltext{2}{Sterrewacht Leiden, Leiden University, NL-2300 RA Leiden, Netherlands}
\altaffiltext{3}{Department of Astronomy, University of California, Berkeley, CA 94720, USA}

\shorttitle{Massive and Newly Dead: High Velocity Dispersions and Strong Balmer Lines at $z\sim1.5$}
\shortauthors{Bezanson et al.}

\newcommand{\unit}[1]{\ensuremath{\, \mathrm{#1}}}

\begin{abstract} 

We present deep Keck/LRIS spectroscopy and HST/WFC3 imaging in the rest-frame
optical for a sample of eight galaxies at $z\sim 1.5$ with high
photometrically-determined stellar
masses. The data
are combined with
VLT/XShooter spectra of five
galaxies from van de Sande et al.\ (2011, 2012 to be submitted). We find that these thirteen
galaxies have high velocity dispersions, with a median of
$\sigma=301\unit{km\,s^{-1}}$. This high value is consistent
with their relatively high stellar masses and compact sizes.
We study their stellar populations using the strength of Balmer
absorption lines, which are not sensitive to dust absorption.
We find a large range in Balmer absorption strength, with many galaxies
showing very strong lines indicating young ages. The median
 $\mathrm{H\delta_A}$ equivalent width, determined directly or inferred from the
H10 line, is 5.4\,\AA, indicating a luminosity-weighted age of $\sim 1$\,Gyr.
Although this value may be biased towards higher values because of selection effects,
high-dispersion galaxies with such young ages are extremely rare in the
local Universe. Interestingly we do not find a simple correlation with
rest-frame $U-V$
color: some of the reddest galaxies have very strong Balmer absorption lines.
These results demonstrate
that many high-dispersion galaxies at $z\sim 1.5$
were quenched recently. This implies that
there must be a population of star-forming progenitors at $z\sim2$
with high velocity dispersions or linewidths, which are notoriously
absent from CO/H$\alpha$ selected surveys.

\end{abstract}

\section{Introduction}

In the local universe, massive galaxies are primarily ellipticals with
little ongoing star-formation and red colors.
Recent surveys  have identified plausible progenitors of these
galaxies at redshifts $z=1-2$, finding that they have small
sizes for their mass \citep[e.g.][]{daddi:05,trujillo:06,toft:07,zirm:07,dokkumnic:08}. Furthermore, by using
rest-frame colors to isolate the bluest dead
galaxies \citep[e.g.][]{kriek:10,whitaker:12a}, it appears that
a significant fraction of them have young ages.
These blue galaxies may be similar to the well-known
``E+A'' galaxies found among low mass galaxies at lower redshift
\citep[e.g.][]{dresslergunn:83,couch:87,zabludoff:96}, although
\citet{whitaker:12a} note that the label ``A galaxies''
may be more appropriate at high redshift
as they may lack an underlying old
component.

These results are quite exciting and suggest that at $z\sim1-2$ we are
beginning to see the build-up of the massive end of the red
sequence. However, this conclusion is based on two main
assumptions. First, the inclusion of these young galaxies in samples
of massive galaxies depends on estimates of stellar mass. Such
estimates suffer from systematic uncertainties such as the IMF and the
contribution of thermally pulsating asymptotic giant branch (TPAGB)
stars \citep[e.g.][]{maraston:05,kriek:10,zibetti:12}. Second,
especially at $z\gtrsim1.5$, the identification of galaxies as young
and old, star-forming vs. quiescent is again based on the SED shapes,
more specifically on the shape of the Balmer/$4000\unit{\AA}$ break,
which depends sensitively on proper treatment of dust reddening and thus is uncertain. Up to now, very few direct measurements of the ages
\citep[e.g.][]{kriek:09} and masses
\citep[e.g.][]{cappellari:09,dokkumnature:09,newman:10,onodera:10,martinezmanso:11,sande:11,toft:12}
of massive galaxies exist at $z>1$.

With spectroscopic data, one can circumvent these
uncertainties.
Balmer absorption lines, which are prominent in the spectra
of A stars and will only be present in a galaxy spectrum for
$\sim1\unit{Gyr}$ after star formation has ceased, can be used as
luminosity-weighted age indicators.
As an example, \citet{leborgne:06}
identified recently quenched $\mathrm{H\delta}$-strong (HDS) galaxies at
$z\sim 1.2$ in the Gemini Deep Deep Survey. These measurements
are particularly valuable when combined with velocity dispersions,
which provide a direct
measurement of the depth of the potential well and
can be combined with structural
parameters to calculate dynamical (rather than photometric) masses.

In this study, we present a sample of eight massive galaxies with deep Keck
I spectroscopy using the LRIS instrument (and five with X-Shooter
spectra from \citet[][2012 to be submitted]{sande:11}) in combination with rest-frame
optical imaging from WFC3 and ACS on HST. Using this exceptional
dataset, we are able to confirm that all of the galaxies in the sample
have high velocity dispersions
(and high dynamical masses). From this unique sample, we conclude that
there is a significant population of recently
quenched galaxies with very high velocity
dispersions at $z\sim1.5$.

\section{Data}

\subsection{High-z Surveys \& Target Selection}

For this project, a sample of massive, bright galaxies was selected
from the Newfirm Medium Band Survey (NMBS) and UKIDSS UDS fields. The
NMBS catalogs include photometry in the COSMOS and AEGIS fields in a
multitude of deep broadband and medium band optical and near-IR
wavebands and includes medium band photometry in the near-IR to sample
the Balmer/$4000\unit{\AA}$ break in galaxies at these redshifts
\citep{whitaker:11}. The UDS field \citep{williams:09} also provides
deep optical and near-IR photometry over a large $0.77\unit{deg^2}$
field. For both survey catalogs, we use the EAzY software \citep{eazy}
to calculate photometric redshifts and InterRest \citep{taylor:09} to
calculate rest-frame colors for all galaxies in the fields. We use
FAST \citep{kriek:09} to estimate stellar masses, assuming
\citet{bc:03} (BC03) stellar population models and a
\citet{chabrier:03} Initial Mass Function (IMF). Primary targets were
selected to have $I\leq24.0$ and $\log M_{\star}\gtrsim11$.

\begin{figure*}[t]
  \centering
  \begin{tabular}{cc}
	\includegraphics[width=0.5\textwidth]{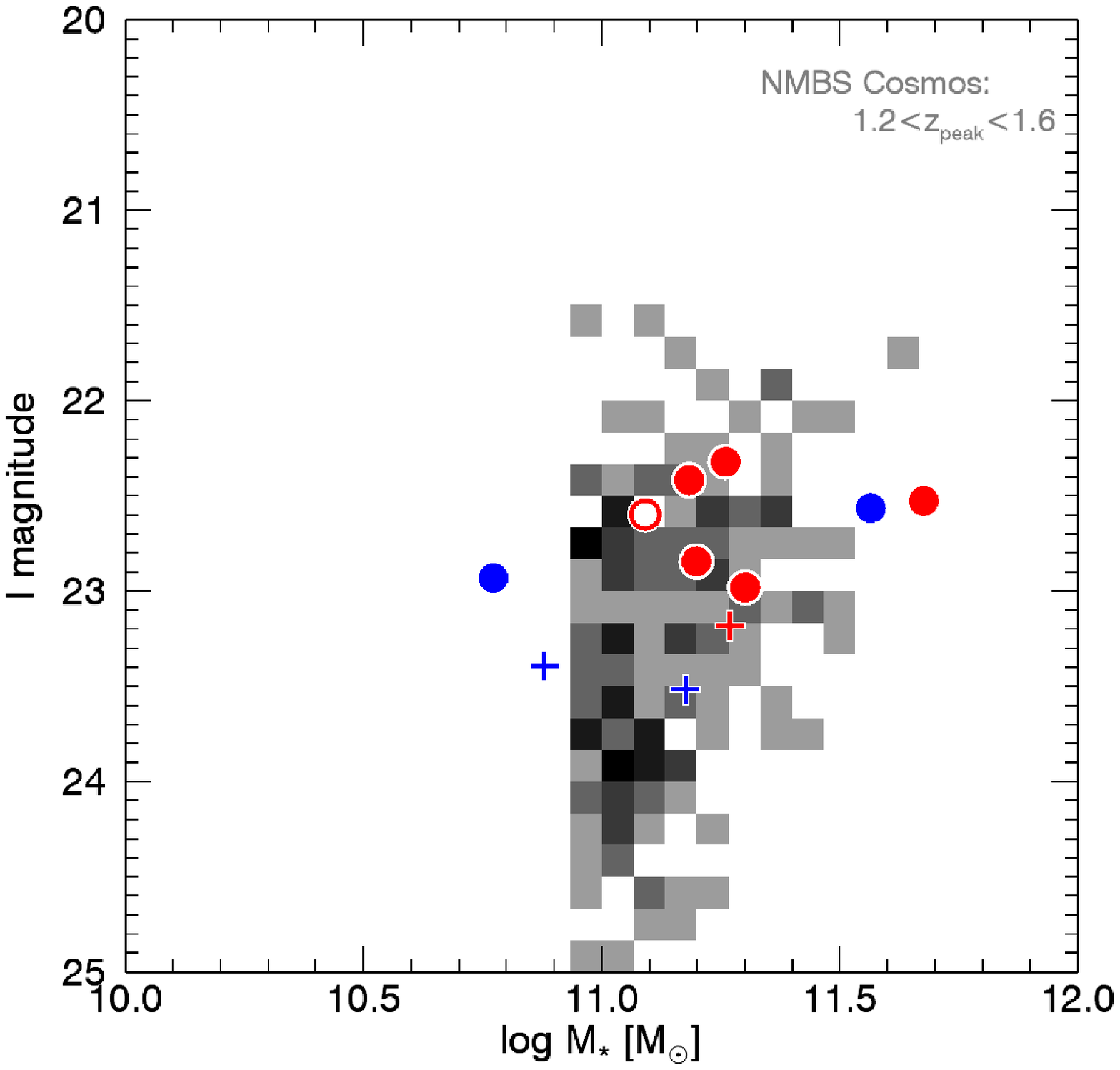} & 
	\includegraphics[width=0.5\textwidth]{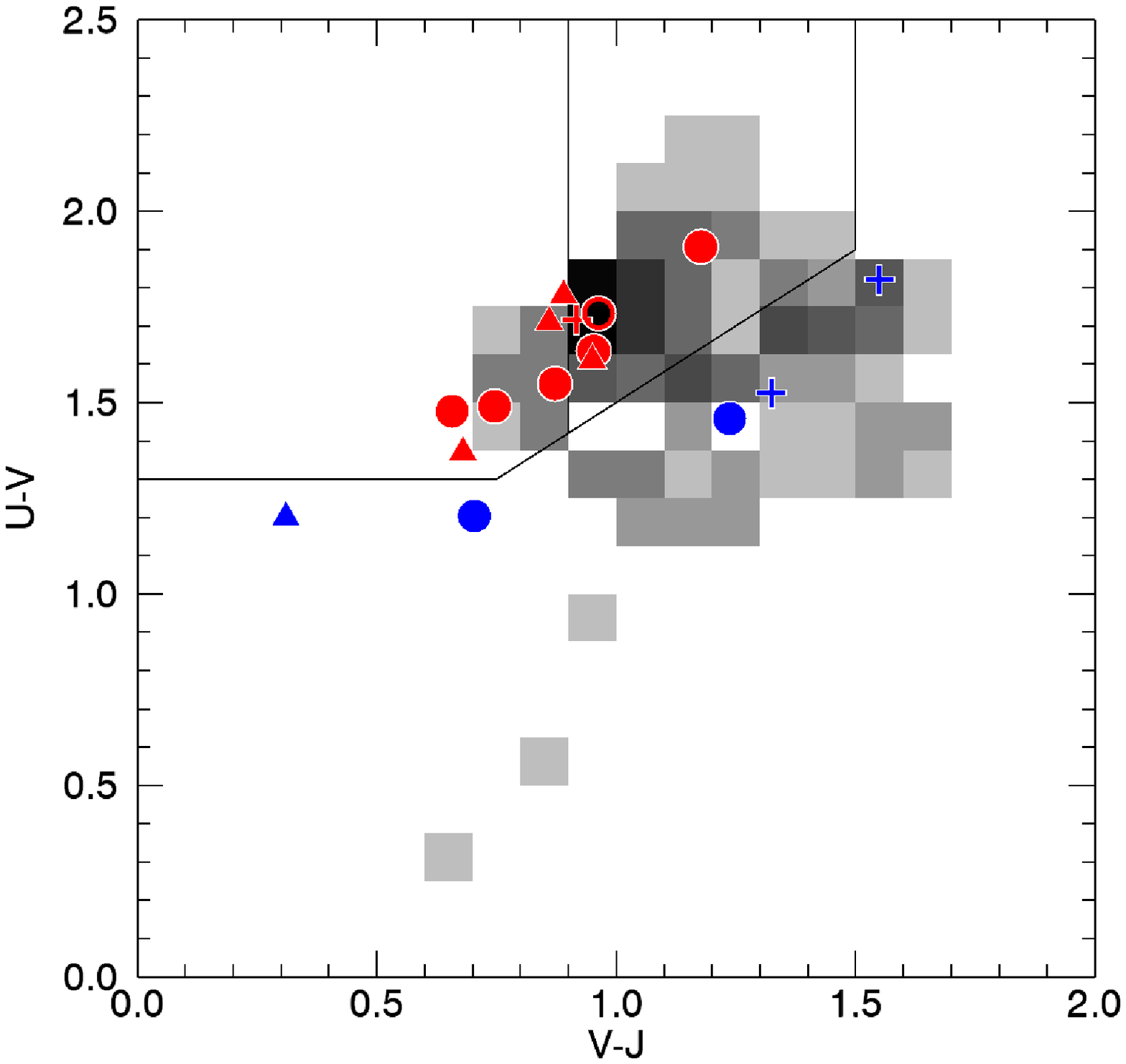} \\
  \end{tabular}
    \caption{Properties of the massive $z\sim1.5$ spectroscopic galaxy sample (colored symbols, crosses indicate insufficient spectroscopic S/N) and overall population of massive $1.2<z<1.6$ galaxies in the NMBS COSMOS field (gray). \emph{(a)} I magnitude vs. stellar mass sample selection. 
    \emph{(b)} rest-frame U-V vs. V-J colors with black outlines highlighting the phase space location of young and old quiescent galaxies as defined in \citet{whitaker:11}. Six galaxies in the sample are red and dead, while two galaxies have ongoing star-formation; their red color can be attributed to dust.  Triangles indicate van de Sande (2011, 2012 to be submitted) sample.}
  \label{fig:select}
 \end{figure*}

In Fig.\ \ref{fig:select}, the properties of massive galaxies
included in our three masks are shown as colored points along with
galaxies with $\log\,M_{\star}>11$ and $1.2<z<1.6$ in the NMBS COSMOS
field. Following \citet[][]{williams:09}
we discriminate between star-forming (blue points) and young/old
quiescent (red points) galaxies using rest-frame $U-V$ and $V-J$
colors.  Our sample spans a large range in $I$-band
magnitudes and rest-frame $U-V$ colors.
However, as is usually the case with spectroscopic samples, our
sample is somewhat biased toward bright magnitudes and
blue colors. This may imply that we are biased towards the inclusion of newly
quenched galaxies, since the bluest, youngest galaxies will also be
the brightest galaxies in the sample and therefore will have the
highest S/N for a given exposure time. Furthermore,
although we do not select explicitly for compactness,
galaxies that have more concentrated light will also have higher
S/N ratios.  Because of this we may also have a bias toward smaller
galaxies at fixed mass (see van de Sande et al, in preparation).

\subsection{Velocity Dispersions and Balmer Line Strengths}

Spectroscopic data were collected using LRIS on the Keck I telescope
during two four-night runs in January and April, 2010. Observations
were performed with the upgraded red arm using the $600\unit{mm^{-1}}$
grating. All observations consist of sequences of three dithered 15
minute exposures, offset by $3"$ each. We observed galaxies in four
different slit masks, two in the COSMOS field
($18\unit{hours},6.75\unit{hours}$), one in the AEGIS field
($15\unit{hours}$) and one in the UDS field
($10.5\unit{hours}$). Reduction of the spectra was performed in IRAF
following standard techniques, as described in
\citet[][]{dokkumfp:03}. No fringe correction was necessary due to the
reduced interference fringing in the LBNL detectors \citep{lris:10}.

\begin{figure*}
	\plotone{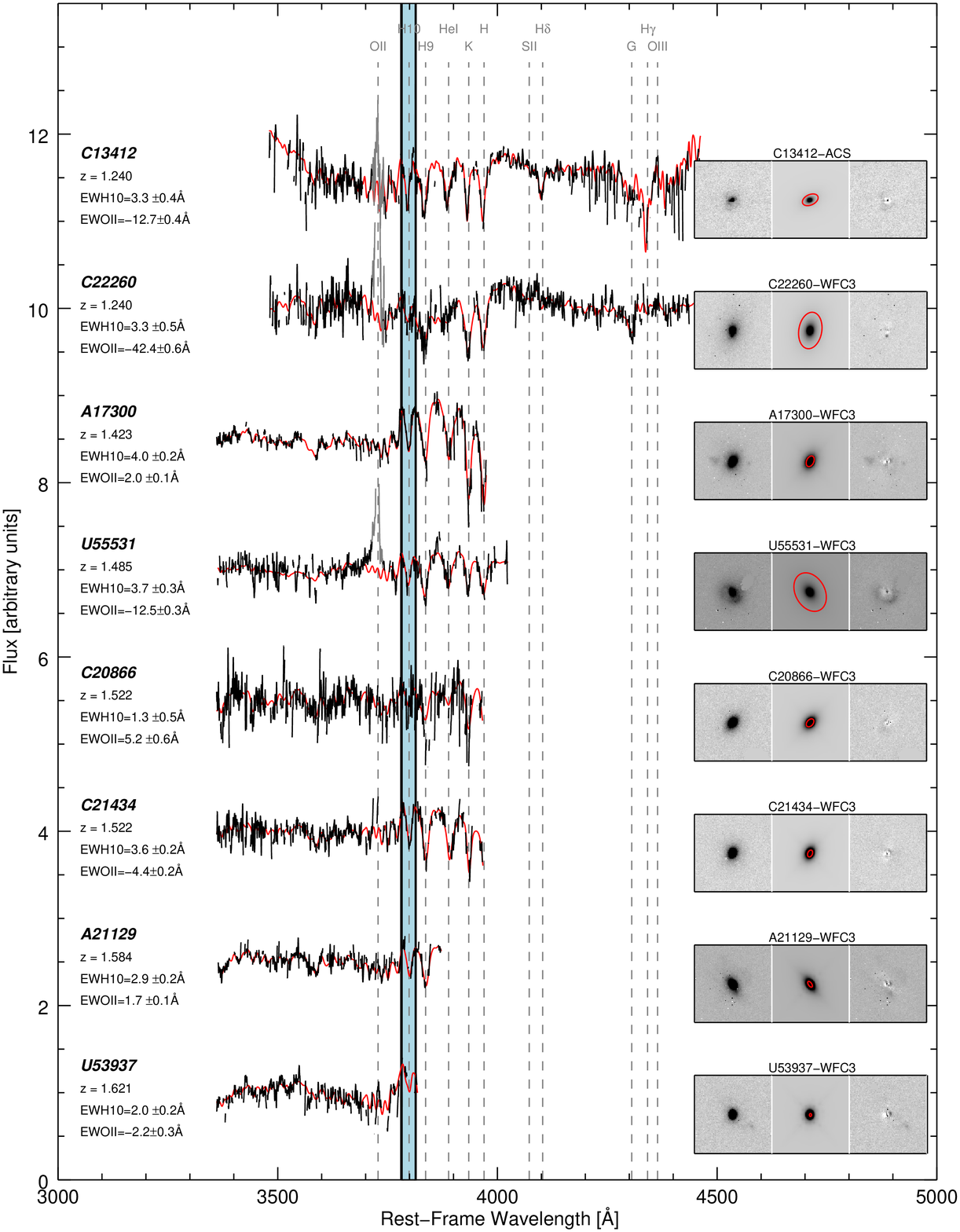}
	\caption{LRIS spectra with best-fit BC03 templates (red) and HST imaging for $z\sim1.5$ massive galaxy sample. $\mathrm{H10}$ bandpass labeled with blue band.  OII emission lines (grey), when present, are masked in dispersion fitting.  Imaging Panels: \emph{left:} Image, \emph{center:} Best-fit S\'ersic model with $r_{\mathrm{e}}$ ellipse (red) \emph{right:} residual image.}
	\label{fig:spectra}
\end{figure*}

We measure velocity dispersions using \emph{PPXF}
\citep{cappellari:04}, allowing the program to fit individual BC03
single stellar population (SSP) models with 8th-order additive and
3rd-order multiplicative polynomials. Best-fit broadened templates
(red) are plotted with galaxy spectra in Fig.\ \ref{fig:spectra}. We
estimate the measurement errors on the dispersion measurements by
shuffling the residuals from the fits, adding them back to the
template and allowing \emph{PPXF} to refit the dispersion.  We only include galaxies for which the measurement error is less than $10\%$ of
the measured velocity dispersion in the following analysis. We verify
that velocity dispersion measurements are not strongly dependent on
the choice of polynomials and are stable within the quoted error bars.
Velocity dispersions are aperture corrected to $r_{\mathrm{e}}/8$ using the size
measurements described in the following subsection and the
prescription $\sigma_0=\sigma_{ap}(8.0r_{ap}/r_{\mathrm{e}})^{0.066}$, based on
the dynamical modeling of the SAURON sample \citep{cappellari:06}.

Traditionally, $\mathrm{H\delta_A}$ has been used as a galaxy age
indicator \citep[e.g.][]{kauffmanncol:03}. However due to the
redshifts of these galaxies and spectral coverage of LRIS, it is only
available for the lowest redshift galaxies in this sample. Instead, we
measure $\mathrm{H10}$; defining bandpasses to measure a
pseudocontinuum ($[3780.0-3785.0\unit{\AA}]$ and
$[3810.0-3815.0\unit{\AA}]$) and line ($[3783.0-3813.0\unit{\AA}]$) from
all spectra, convolved to $300\unit{km\,s^{-1}}$. Template fluxes from the
dynamical modeling are substituted for masked pixels and errors are
estimated using 1000 bootstrap iterations with shuffled residuals. We
convert from $\mathrm{H10}$ to Lick $\mathrm{H\delta_A}$ as
follows. Using BC03 templates, also convolved to
$\sigma=300\unit{km\,s^{-1}}$, we measure both $\mathrm{H10}$ and
$\mathrm{H\delta_A}$ and find a tight correlation for templates older
than $\sim100\unit{Myr}$. We fit the relation with a polynomial and
calculate inferred $\mathrm{H\delta_A}$ as:

\begin{equation}
\mathrm{H\delta_A}=-16.37+10.45\mathrm{H10}-1.44\mathrm{H10}^2+0.08\mathrm{H10}^3.
\end{equation}

We also measure $3727\unit{\AA}$ OII emission using the bandpasses
defined by \cite{fisher:98} and estimate errors as described above.

\subsection{Structural Parameters}

Sizes of all galaxies in our sample are measured from Hubble
Space Telescope (HST) imaging. For seven of the galaxies, we use F160W
imaging from the  WFC3 camera, sampling the rest-frame
$R$ band. Observations were conducted with a four
point dither pattern, offset by half pixel increments. We interlace
the flat-fielded frames to increase the resolution of the galaxy
images and better sample the point-spread function (PSF), without
introducing smoothing inherent in drizzling. We use a similarly
interlaced TinyTim PSF, initially oversampled by a factor of 10,
shifted to reflect the dither pattern, interlaced in the same manner
and finally binned to the resolution of the interlaced galaxy
images. We extract postage stamps from the interlaced frames centered
on each galaxy, aggressively mask other objects using SExtractor
\citep{bertin:96} segmentation maps and use Galfit \citep{galfit} to
fit S\'{e}rsic profiles to the galaxy images. We verify that these
measurements are quite consistent (with a scatter of 3\%) with those
made from the drizzled images produced by the HST pipeline and a
nearby PSF star.

For the single galaxy for which WFC3 imaging is not
available we measure its size from ACS imaging and distinguish this
data point as an open circle in Figures \ref{fig:properties1} and
\ref{fig:properties2}. We use ACS imaging from the COSMOS team,
cutting a postage stamp from the v.1.3 ACS F814W mosaic from
\citet{scoville:07} and using an isolated PSF star. We note that
although the F814W is blueward of rest-frame optical for this galaxy,
it is at the lowest redshift of the sample and the bandpass effects
should be minimal.

\begin{deluxetable*}{cccccccccccc}
\tablecaption{Measurements for $z\sim1.5$ massive galaxy sample with LRIS spectroscopy.}
\tablehead{
\colhead{ID} & \colhead{$z_{\mathrm{spec}}$} & \colhead{$r_{\mathrm{e,c}}$} & \colhead{n} & \colhead{b/a} & \colhead{$M_{\star}$} & \colhead{$\sigma_{\mathrm{inf}}$} & \colhead{$\sigma_{\mathrm{meas}}$} & \colhead{$\sigma_{0}$} & \colhead{$M_{\mathrm{dyn}}$} & \colhead{H10} & \colhead{OII} \\
\colhead{} & \colhead{} & \colhead{$\unit{[kpc]}$} & \colhead{} & \colhead{} & \colhead{$\unit{[M_{\odot}}]$} & \colhead{$\unit{[km\,s^{-1}]}$} & \colhead{$\unit{[km\,s^{-1}]}$} & \colhead{$\unit{[km\,s^{-1}]}$} & \colhead{$\unit{[M_{\odot}]}$} & \colhead{$\unit{[\AA]}$} & \colhead{$\unit{[\AA]}$}}

\startdata
$A17300$ & $1.4235$ & $2.9$& $5.3$ & $0.64$ & $11.26\pm0.1$ & $375$ & $265\pm7$ & $312\pm8$ & $11.36\pm0.05$ & $4.0\pm0.2$ & $2.0\pm0.1$ \\[+1.5ex] 
$A21129$ & $1.5839$ & $1.5$ & $5.0$ & $0.50$ & $11.18\pm0.1$ & $459$ & $260\pm9$ & $319\pm12$ & $11.12\pm0.05$ & $2.9\pm0.2$ & $1.7\pm0.1$ \\[+1.5ex] 
$C13412\footnote[1]{Morphology measured from ACS imaging}$ & $1.2395$ & $4.2$ & $6.0$ & $0.62$ & $11.09\pm0.1$ & $274$ & $116\pm8$ & $133\pm9$ & $10.72\pm0.06$ & $3.3\pm0.3$ & $-12.7\pm0.4$ \\[+1.5ex] 
$C22260$ & $1.2399$ & $9.3$ & $4.1$ & $0.63$ & $11.68\pm0.1$ & $292$ & $249\pm16$ & $271\pm17$ & $11.86\pm0.06$ & $3.3\pm0.5$ & $-42.4\pm0.7$ \\[+1.5ex] 
$C21434$ & $1.5223$ & $1.9$ & $3.1$ & $0.65$ & $11.20\pm0.1$ & $337$ & $218\pm16$ & $264\pm20$ & $11.24\pm0.06$ & $3.6\pm0.2$ & $-4.4\pm0.3$ \\[+1.5ex] 
$C20866$ & $1.5222$ & $2.4$ & $3.0$ & $0.67$ & $11.30\pm0.1$ & $329$ & $272\pm23$ & $324\pm27$ & $11.54\pm0.07$ & $1.3\pm0.6$ & $5.2\pm0.6$ \\[+1.5ex] 
$U53937$ & $1.6210$ & $0.8$ & $3.6$ & $0.75$ & $10.77\pm0.1$ & $335$ & $231\pm19$ & $296\pm24$ & $10.92\pm0.07$ & $2.0\pm0.2$ & $-2.2\pm0.3$ \\[+1.5ex] 
$U55531$ & $1.4848$ & $11.2$ & $4.9$ & $0.72$ & $11.57\pm0.1$ & $261$ & $257\pm24$ & $277\pm26$ & $11.87\pm0.07$ & $3.7\pm0.3$ & $-12.5\pm0.3$ \\ 
\enddata
\end{deluxetable*}

\section{The Extreme Dynamics and Stellar Populations of $z\sim1.5$ Massive Galaxies}

\subsection{High Velocity Dispersions and Masses}

\begin{figure*}[]
  \centering
  \begin{tabular}{ll}
	\includegraphics[width=0.5\textwidth]{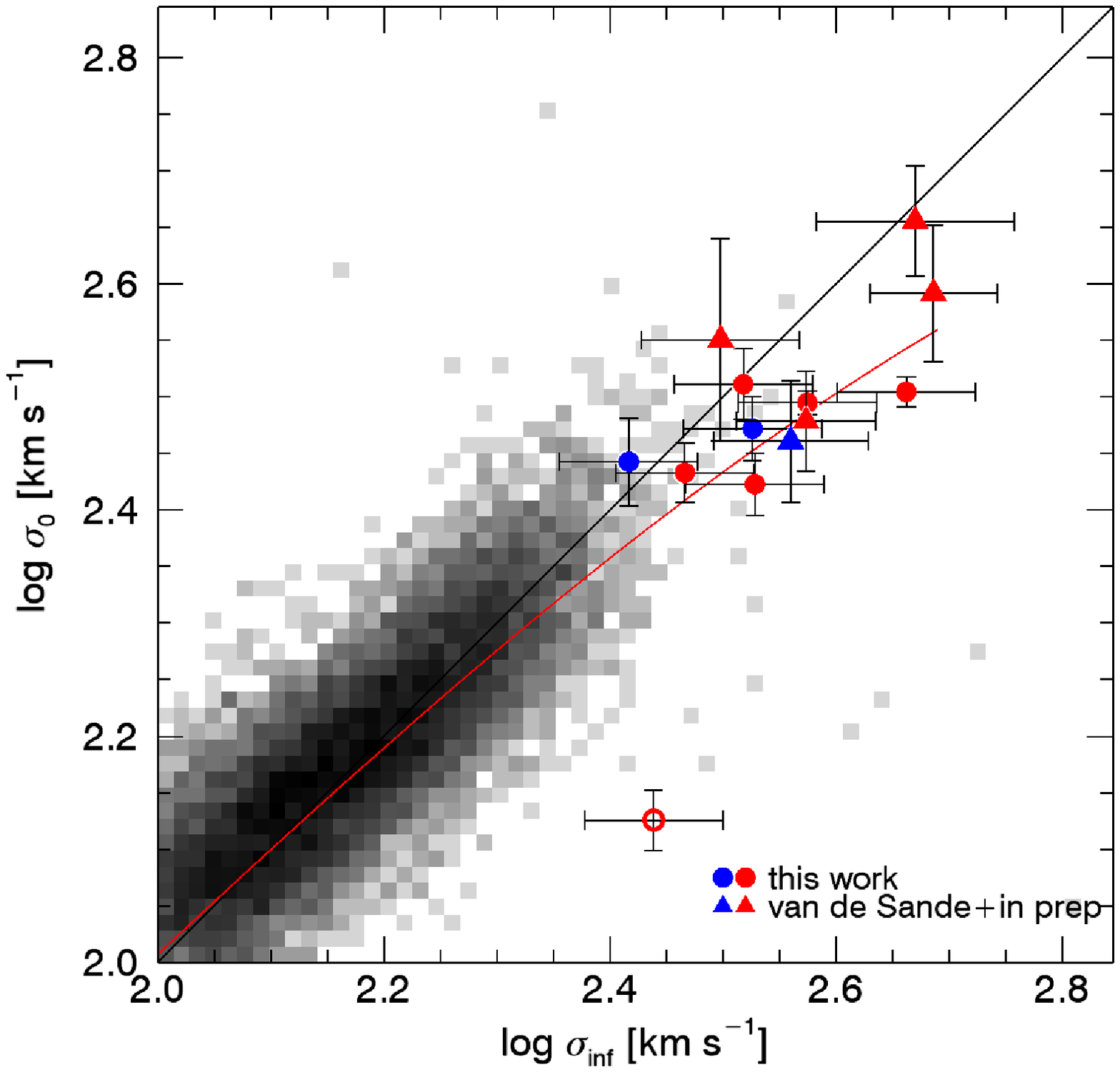} &
	\includegraphics[width=0.5\textwidth]{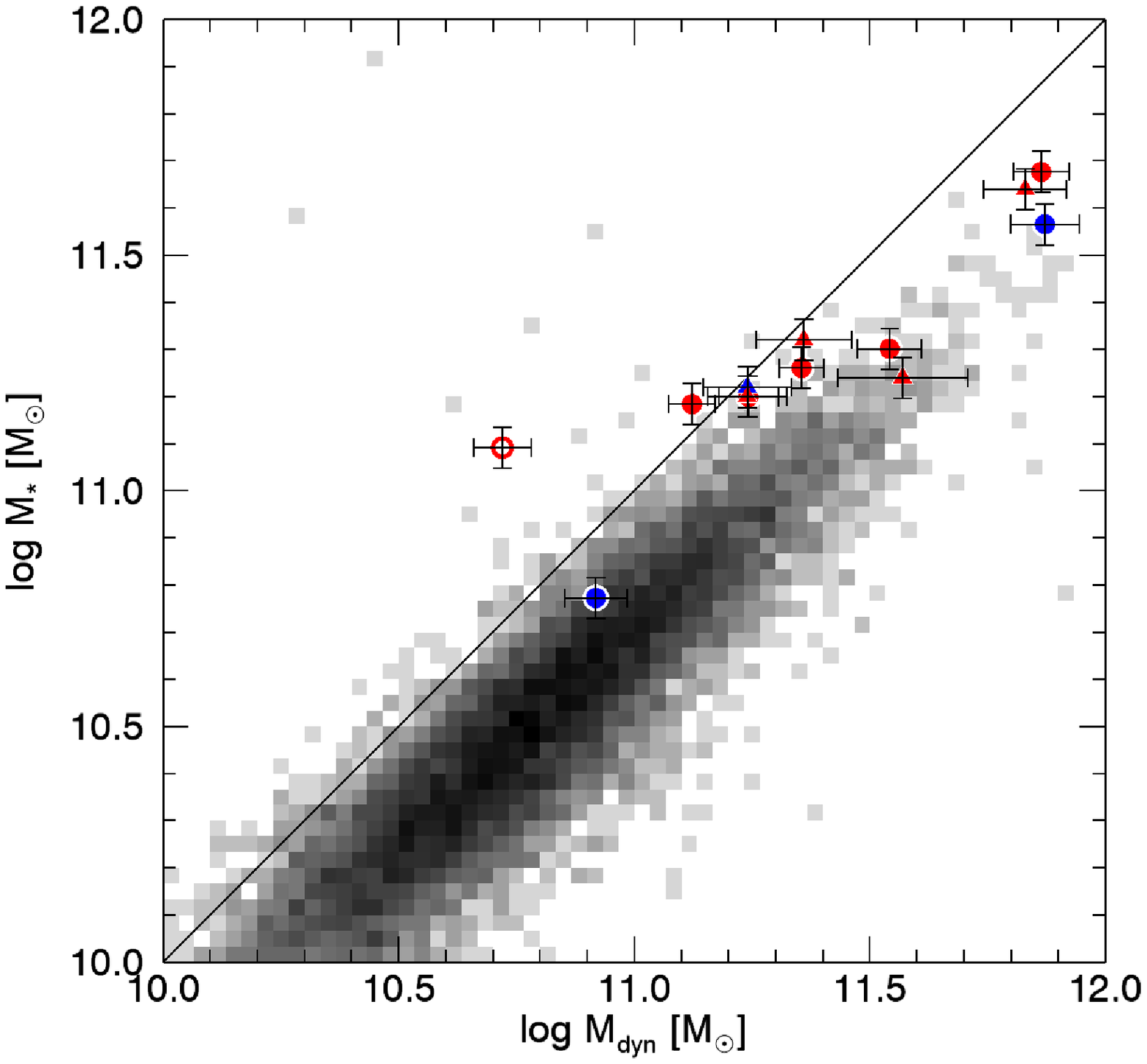} \\
  \end{tabular}
    \caption{\emph{(a)} Inferred versus measured velocity dispersions for $z\sim1.5$ sample (colored points) and massive galaxy sample in the SDSS (greyscale). LRIS sample are circles (WFC3-filled, ACS-open) and X-Shooter sample (van de Sande, in prep) are triangles. As in Fig.\ \ref{fig:select}, color-coding is based on rest-frame color derived quiescence. The high redshift galaxies have very high dispersions, coinciding with the tail of the SDSS distribution. Inferred dispersion appears to be a reasonable predictor of intrinsic velocity dispersion, even at the highest dispersions. \emph{(b)} Dynamical versus stellar mass for high and low redshift galaxies.}
  \label{fig:properties1}
 \end{figure*}

As can be seen in Table 1, the galaxies have very high velocity dispersions; the median $\sigma_0$ for the complete sample is $301\unit{km\,s^{-1}}$. In the local Universe,
such high dispersions are rare, and typical of giant galaxies
at the centers of groups or clusters such as NGC 4374,
NGC 4472, and M87. This is demonstrated in the left
panel of Fig.\ \ref{fig:properties1}, where
we show our sample with colored points and data from the SDSS in
grey. The $z\sim 1.5$ galaxies fall on the high dispersion
tail of the SDSS distribution.

The high dispersions are fully consistent with the high stellar masses and small sizes
of the galaxies. This is shown explicitly in Fig.\ \ref{fig:properties1}a,
as we compare the measured velocity dispersions to the velocity dispersion
inferred from the size, Sersic index, and stellar mass, scaling by the
average factor between stellar mass and structural-corrected $M_{\mathrm{dyn}}$
in the SDSS \citep{franx:08,taylor:09,sande:11,bezanson:11,bezanson:12a}. We
calculate $\sigma_{\mathrm{inf}}$ as:
\begin{equation}
\sigma_{\mathrm{inf}}=\sqrt{\frac{GM_{\star}}{0.557K_{\mathrm{v}}(n)r_{\mathrm{e}}}},
\label{eqn:siginf}
\end{equation}
with
\begin{equation}
K_{\mathrm{V}}(n)=\frac{73.32}{10.465+(n-0.94)^2}+0.954,
\end{equation}
as derived by \citet{bertin:02}.  This constant corresponds
to velocity dispersions within $r_{\mathrm{e}}/8$.

There is good agreement between inferred and measured velocity dispersions, although there may be a small systematic
offset. This may be caused by the fact that
galaxies with extremely high velocity dispersions are rare, locally
and $z\sim1.5$ \citep[e.g.][]{bezanson:11}. Based on the steep high dispersion tail of the
velocity dispersion function (VDF), galaxies with high $\sigma_{\mathrm{inf}}$
are more likely to be scattered above the $\sigma_{\mathrm{inf}}-\sigma_0$
relation than to have intrinsically high $\sigma_0$. Therefore for a sample of galaxies drawn from this distribution, the relation will
veer from one-to-one at high dispersions. We simulate this
effect by selecting samples of galaxies uniformly in $\sigma_{\mathrm{inf}}$ from
the local $\Phi(\sigma_0)$  \citep{bernardi:10}
plus $0.06\unit{dex}$ scatter between $\sigma_0$ and
$\sigma_{\mathrm{inf}}$. We include a polynomial fit to the resulting relation (red line) in Fig.\ \ref{fig:properties1}a. Inferred
dispersion remains a good predictor of measured velocity dispersion
for this sample, although the data suggest that the scatter may be
higher at $z\sim1.5$ \citep{bezanson:11}.

For completeness, we directly compare stellar and dynamical masses in Fig.\ \ref{fig:properties1}b. Dynamical masses were calculated
for each galaxy from its size, velocity
dispersion and best-fit S\'ersic index
\citep[e.g.][]{bertin:02,cappellari:06,taylor:10}:
\begin{equation}
M_{\mathrm{dyn}}=\frac{K_{\mathrm{V}}(n)\sigma^2\,r_{\mathrm{e}}}{G}.
\end{equation}
The most massive galaxies in the sample lie directly on top of the local
relation, while the lower mass galaxies lie closer to the one-to-one
line. There is a hint that overall stellar masses are closer to dynamical
masses at higher redshifts, increasing the ratio from a median of
$\log(M_{\star}/M_{\mathrm{dyn}})=0.27\unit{dex}$ locally to $0.09\unit{dex}$
at $z\sim1.5$. However, the local stellar-to-dynamical mass relation
exhibits a fair amount of scatter, and due to the biased nature of this
sample of galaxies it is hard to disentangle whether this offset is
due to evolution and/or selection effects. In all cases, we verify that
the galaxies in this sample are dynamically massive.

\begin{figure*}[!th]
  \centering
  \begin{tabular}{cc}
	\includegraphics[scale=0.35]{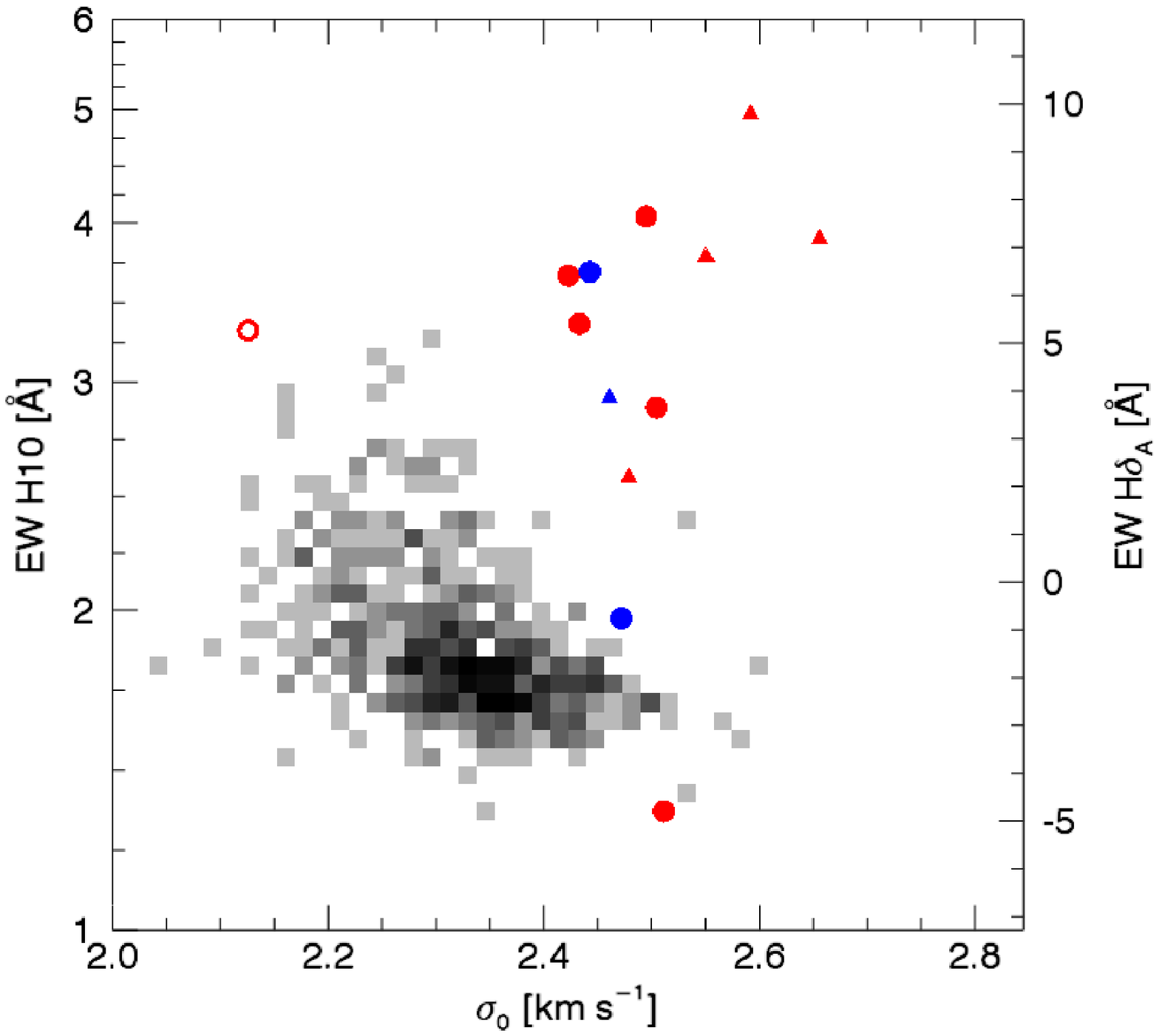} &
  	\includegraphics[scale=0.35]{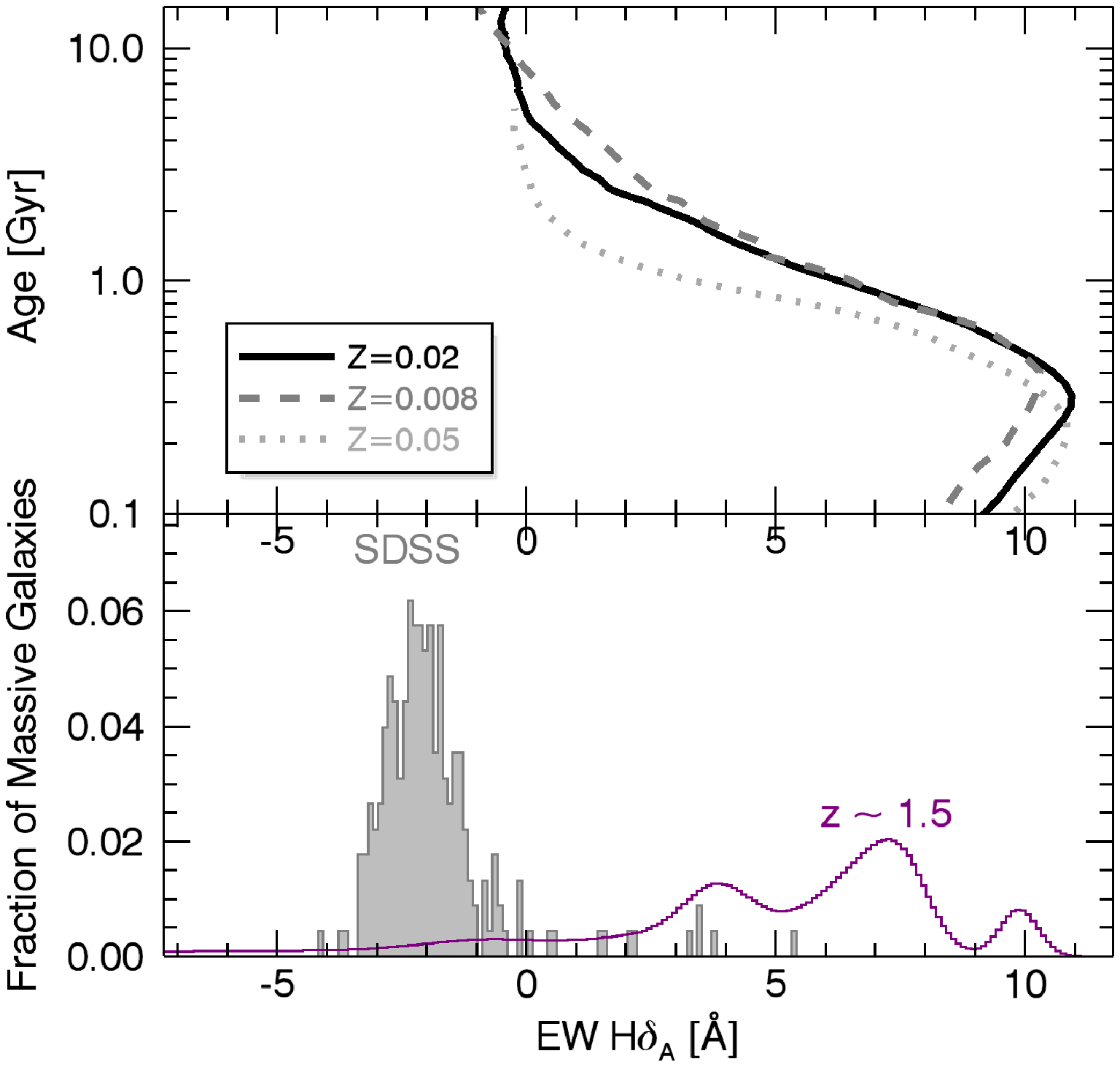} \\
\end{tabular}
\begin{tabular}{c}
  	\includegraphics[scale=0.35]{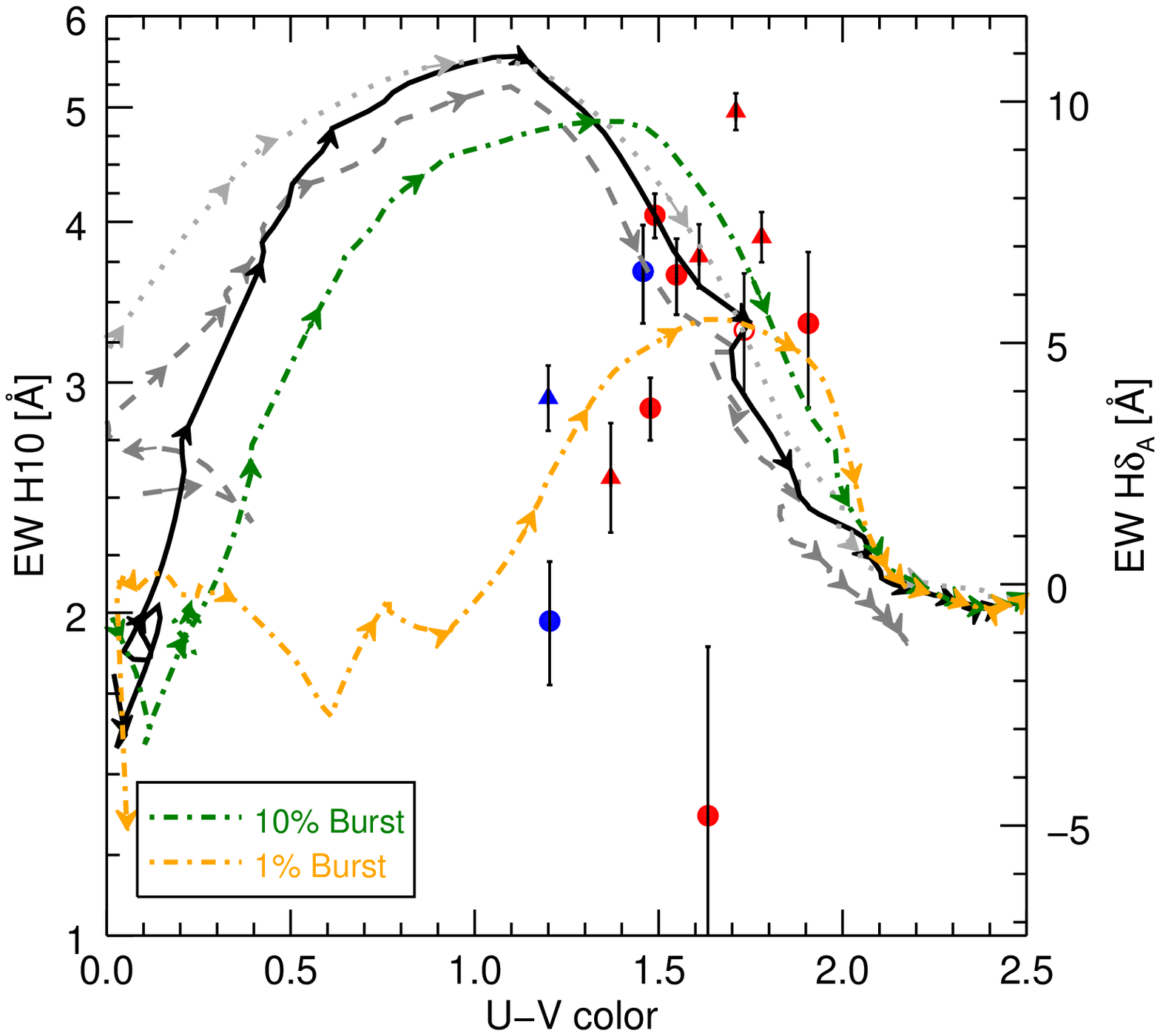} \\
  \end{tabular}
    \caption{\emph{(a)} Balmer EWs versus velocity dispersion for massive galaxies. The majority of galaxies in this sample have the highest velocity dispersions known at any redshift. Unlike the uniformly old, dead stellar populations in comparable galaxies at $z\sim0$, many of these galaxies have strong Balmer lines, implying recent quenching.\emph{(b)} Top: Age versus $\mathrm{H\delta_A}$ for BC03 SSP templates of varied metallicity. Bottom: $\mathrm{H\delta_A}$ distribution from high-dispersion ($\log\sigma>2.4$) galaxies in the SDSS (grey) and $z\sim1.5$ distribution (purple) with individual galaxies included as gaussians of width equal to their measurement errors. In striking contrast with local high-dispersion galaxies, which have uniformly weak Balmer lines and old luminosity-weighted ages, this sample exhibits a broad distribution of EWs including many with strong Balmer lines and young ages.  \emph{(c)} Balmer EWs versus rest-frame U-V colors relative to BC03 SSP (grey) and double-burst (green and orange) models.  Even though the colors and Balmer lines suggest a similar range in ages for the galaxies in this sample, the galaxies do not follow the simple model tracks which suggests the importance of dust or secondary bursts of star formation in these systems.}
  \label{fig:properties2}
 \end{figure*}
 
\subsection{Using Balmer Line Strengths to Identify Recently Quenched Galaxies}

We find that the galaxies exhibit a large range in Balmer line strengths,
with $\mathrm{H10}$ ranging from
$1.3-4.0\unit{\AA}\,(\mathrm{H\delta_A}=-4.8-7.6\unit{\AA}$).
The immediate implication is that they have a large range in
ages and/or a recent burst contribution. Furthermore, the median Balmer line strength is quite
high at $\mathrm{H10}=3.3\unit{\AA}\,(\mathrm{H\delta_A}=5.4\unit{\AA})$.
This is a remarkable result given the high dispersions of the
galaxies. Fig.\ \ref{fig:properties2}a shows the relation between
Balmer line strength and velocity dispersion for SDSS galaxies
(grey) and our data (colored). The Balmer lines at $z\sim 1.5$
are dramatically stronger, and show a much larger range, than at
$z=0$. This is arguably the most direct evidence yet for rapid
quenching of massive galaxies at $z\sim 1.5$, as the quantities
on both axes are directly measured and completely
independent from photometric data.
We note that
\citet{leborgne:06} also showed a strongly increasing fraction of
galaxies with strong Balmer lines out to $z\sim1.2$, although
their sample had lower photometric masses
($\log\,M>10.2$) and lacked velocity dispersion measurements.

In Fig.\ \ref{fig:properties2}b we interpret the Balmer lines in the
context of the stellar population synthesis (SPS) models of Bruzual
\& Charlot (2003). The distribution
of Balmer line strengths is indicated in the bottom panel and
can be interpreted in terms of ages using the top panel.
The observed range in line strenghts implies luminosity-weighted
ages ranging from several hundred Myrs to maximally old stellar
populations, with a median of $\sim1\unit{Gyr}$. 

An obvious question is whether we could have drawn the same
conclusions from photometric data. 
Fig.\ \ref{fig:properties2}c compares the Balmer line strengths and rest-frame
$U-V$ colors for the galaxies in this sample and to SPS models.
The models are BC03 single population tracks 
for three different metallicities (grey) and double-burst models
(green and orange dot-dashed) in which $10\%$ and $1\%$ of the galaxy's mass is formed in a
secondary burst $4\unit{Gyr}$ after the original episode of
star-formation. Although the {\em range} in colors of this sample of galaxies implies
a similar range in ages as the Balmer lines do, the data do not
follow the model predictions. Specifically, the data show no anti-correlation
between color and Balmer line strength, as would be expected.
Some of the bluest galaxies have weak Balmer lines, and some of the
reddest have strong Balmer lines.

A possible explanation is that the colors are affected by dust reddening,
and Fig.\ \ref{fig:properties2}c illustrates the importance of independent
measures of age. 
To compare the intrinsic distribution of colors/ages, one
could design ``dust-corrected" colors \citep[e.g.][]{brammer:09},
however these depend strongly on SED modeling and well-determined dust
estimates. We also note that the sample of 13 is barely adequate for
this particular test; larger, unbiased
samples may show the expected correlation
between Balmer lines and colors, as strongly suggested by the complete
samples of Whitaker et al.\ (2012).

\section{Discussion and Conclusions}

We have identified a population of galaxies at $z\sim 1.5$
with high velocity dispersions and strong Balmer absorption lines.
These results confirm that by $z\sim1.5$ the population of massive
galaxies is much less uniform than it is locally: instead of being
red, dead ellipticals, many have experienced recent star-formation
within $\sim\,1\unit{Gyr}$ \citep[see
  also][]{leborgne:06,dokkumbrammer:10,dokkum:11,whitaker:12a}.
A direct implication is
that $\lesssim1\unit{Gyr}$ earlier, at $z\sim2$, there must
exist highly star-forming progenitor galaxies with similarly high
velocity dispersions. Such galaxies have yet to be identified,
and the fact that some of the youngest galaxies in our sample
are also the reddest suggests that these progenitors may have a very
high dust content.

The results are also consistent with our study of the velocity
dispersion function of quiescent and star forming galaxies
\citep{bezanson:12a}, as we concluded that
quenching must occur
efficiently for galaxies with high velocity dispersions.
This conclusion was based on the
rapid build-up of the VDF of quenched galaxies and the stability of
the star-forming VDF since $z\sim1.5$. The discovery of this
population of young, high-dispersion galaxies is strong evidence in
favor of this model.

In addition to having different stellar populations at $z\gtrsim1.5$,
a flurry of studies in the last decade have uncovered the dramatic
size evolution that massive galaxies have undergone in the last 10
billion years
\citep[e.g.][]{daddi:05,trujillo:06,toft:07,zirm:07,dokkumnic:08}. The
temporal correlation between the structural evolution and recently
quenched stars presents a tempting connection between the events which
halt star-formation in massive galaxies and at the same time cause
them to be compact relative to their local
analogs. \citet{whitaker:12a} found there to be no correlation between
color of quiescent galaxies and their sizes, however it would be
interesting to investigate the size evolution based on a larger, more
representative sample of spectroscopic data to estimate galaxy ages.

We note that the selection and observational biases may influence
the distribution of galaxies in this sample. In this context our study
is complimentary to \citet{whitaker:12a}, which is mass complete for
galaxies in the NMBS but relies solely on SEDs to estimate photometric
redshifts, stellar masses and rest-frame colors to derive galaxy
ages. However, we note that while qualitatively
the results are similar, we find a puzzling lack of (anti-)correlation
between color and Balmer line strength:
neither the star-forming galaxies nor the bluest quiescent
galaxies in our sample necessarily have the strongest Balmer
lines.

As is often the case, larger samples are needed to explore these
issues. With increasingly powerful multiplexed spectroscopic
instruments and excellent space-based imaging we can construct large
samples of galaxies at this pivotal epoch and simultaneously study the
structural properties and stellar populations of all types of
dynamically massive galaxies.


\end{document}